\documentclass[letterpaper,usenatbib]{mn2e}
\usepackage{amsmath}
\usepackage[breaklinks, colorlinks, citecolor=blue]{hyperref}
\usepackage{epsfig}
\usepackage{multirow}
\usepackage{times}
\usepackage{verbatim}
\def\gtsim{~\rlap{$>$}{\lower 1.0ex\hbox{$\sim$}}}
\def\ltsim{~\rlap{$<$}{\lower 1.0ex\hbox{$\sim$}}}

\title[Star formation in NGC 5253]{Tests of star formation metrics in the low metallicity galaxy NGC~5253 using ALMA observations of H30$\alpha$ line emission}

\author[G. J. Bendo et al.]
    {G. J. Bendo$^{1,2}$, R. E. Miura$^{3}$, D. Espada$^3$, 
      K. Nakanishi$^{3,4}$, R. J. Beswick$^1$,\newauthor 
      M. J. D'Cruze$^1$, C. Dickinson$^1$, G. A. Fuller$^{1,2}$ \\
    $^1$   Jodrell Bank Centre for Astrophysics,
           School of Physics and Astronomy, The University of Manchester, 
           Oxford Road,\\ Manchester M13 9PL, United Kingdom\\
    $^2$   UK ALMA Regional Centre Node\\
    $^3$   National Astronomical Observatory of Japan, 
           2-21-1 Osawa, Mitaka, Tokyo, 181-8588, Japan\\
    $^4$   The Graduate University for Advanced Studies (Sokendai), 
           2-21-1 Osawa, Mitaka, Tokyo 181-0015, Japan    
}
\date{}
\pagerange{\pageref{firstpage}--\pageref{lastpage}}
\pubyear{}

\begin{document}
\label{firstpage}
\maketitle

\begin{abstract}
We use Atacama Large Millimeter/submillimeter Array (ALMA) observations of H30$\alpha$ (231.90~GHz) emission from the low metallicity dwarf galaxy NGC~5253 to measure the star formation rate (SFR) within the galaxy and to test the reliability of SFRs derived from other commonly-used metrics.  The H30$\alpha$ emission, which originates mainly from the central starburst, yields a photoionizing photon production rate of (1.9$\pm$0.3)$\times$$10^{52}$~s$^{-1}$ and an SFR of 0.087$\pm$0.013 M$_\odot$ yr$^{-1}$ based on conversions that account for the low metallicity of the galaxy and for stellar rotation.  Among the other star formation metrics we examined, the SFR calculated from the total infrared flux was statistically equivalent to the values from the H30$\alpha$ data.  The SFR based on previously-published versions of the H$\alpha$ flux that were extinction corrected using Pa$\alpha$ and Pa$\beta$ lines were lower than but also statistically similar to the H30$\alpha$ value.  The mid-infrared (22~$\mu$m) flux density and the composite star formation tracer based on H$\alpha$ and mid-infrared emission give SFRs that were significantly higher because the dust emission appears unusually hot compared to typical spiral galaxies.  Conversely, the 70 and 160~$\mu$m flux densities yielded SFR lower than the H30$\alpha$ value, although the SFRs from the 70~$\mu$m and H30$\alpha$ data were within 1-2$\sigma$ of each other.  While further analysis on a broader range of galaxies are needed, these results are instructive of the best and worst methods to use when measuring SFR in low metallicity dwarf galaxies like NGC~5253.
\end{abstract}


\section{Introduction}
\label{s_intro}
\addtocounter{footnote}{4}

Star formation in other galaxies is typically identified by looking at tracers of young stellar populations, including either photoionizing stars, ultraviolet-luminous stars, and supernovae.  The most commonly-used star formation tracers are ultraviolet continuum emission; H$\alpha$ (6563~\AA) and other optical and near-infrared recombination lines; mid- and far-infrared continuum emission; and radio continuum emission.  However, each of these tracers have disadvantages when used to measure star formation rates (SFRs).  Ultraviolet continuum and optical recombination line emission directly trace the young stellar populations, but dust obscuration typically affects the SFRs from these tracers.  Near-infrared recombination line emission is less affected by dust obscuration, but it is still a concern in very dusty starburst galaxies.  Dust continuum emission in the infrared is unaffected by dust obscuration except in extreme cases, but since this emission is actually a tracer of bolometric stellar luminosity and not just the younger stellar population, it may yield  an overestimate of the SFR if many evolved stars are present.  Radio continuum emission traces a combination of free-free continuum emission from photoionized gas and synchrotron emission from supernova remnants, so proper spectral decomposition is needed to accurately convert radio emission to SFR. Additionally, the cosmic rays that produce synchrotron emission will travel significant distances through the ISM, making radio emission appear diffused relative to star formation on scales of $\sim$100~pc \citep{murphy06a, murphy06b}.

Higher-order recombination line emission at millimetre and submillimetre wavelengths, which is produced by the same photoionized gas that produce H$\alpha$ and other optical and near-infrared recombination lines, can also be used to measure SFRs.  Unlike ultraviolet, optical, and near-infrared star formation tracers, these millimetre and submillimetre recombination lines are not affected by dust extinction, but unlike infrared and radio synchrotron emission, the recombination lines directly trace the photoionizing stars.  Recombination line emission can also be observed at centimetre and longer wavelengths, but the line emission at these longer wavelengths is generally affected by a combination of masing effects and opacity issues in the photoionized gas, while the millimetre and submillimetre lines are not \citep{gordon90}.

The primary reason why these millimetre and submillimetre recombination lines are not used more frequently as star formation tracers is because the emission is very faint.  Before the Atacama Large Millimeter/submillimeter Array (ALMA) became operational, millimetre and submillimetre recombination line emission has only been detected within three other galaxies: M82 \citep{seaquist94, seaquist96}, NGC 253 \citep{puxley97}, and Arp 220 \citep{anantharamaiah00}.  ALMA, however, has the sensitivity to potentially detect this emission in many more nearby galaxies, including many nearby luminous infrared galaxies and other starbursts \citep{scoville13}.   Detections of recombination line emission from the first three cycles of ALMA observations include measurements of the H40$\alpha$ (99.02~GHz) emission from the centre of NGC~253 \citep{meier15, bendo15b} and H42$\alpha$ (85.69~GHz) emission from the centre of NGC~4945 \citep{bendo16} as well as a marginal detection of H26$\alpha$ (353.62~GHz) in Arp 220 \citep{scoville15}.

The NGC 253 and 4945 analyses included comparisons of SFRs from ALMA recombination line and free-free emission to other star formation tracers in radio and infrared emission.  The results illustrated some of the challenges in measuring star formation rates in other wavebands.  SFRs from radio continuum emission calculated using one of the conversions given by \citet{condon92} or \citet{murphy11} yielded results that differed significantly from the ALMA data.  Recombination line emission at centimetre or longer wavelengths often produced much less accurate results.  For NGC~253 specifically, some of the SFRs from recombination line emission at these longer wavelengths were $\sim$3$\times$ lower than what was determined from the ALMA data, which indicated that the longer wavelength lines may be affected by gas opacity effects.  The SFR from the mid-infrared data for the central starburst in NGC~4945 was $\sim$10$\times$ lower than the SFR from the H42$\alpha$ or free-free emission, but the SFR from the total infrared flux was consistent with the ALMA-based SFRs.  This suggests that the centre of NGC~4945 is so dusty that even the mid-infrared light from the dust itself is heavily obscured.  

While the results for NGC 253 and 4945 have revealed some of the limitations of other star formation tracers, the analyses have mainly focused on radio or infrared data.  The dust extinction is so high in the central starbursts in both galaxies that comparisons of millimetre recombination line emission to ultraviolet or H$\alpha$ emission is not worthwhile.  Such comparisons need to be performed using a less dusty object.

NGC 5253 is a nearby blue compact dwarf galaxy within the M83/Centaurus A Group \citep{karachentsev07} that hosts a starburst nucleus.  Most published distances for the galaxy range from 3 to 4~Mpc; we use a distance of 3.15$\pm$0.20~Mpc \citep{freedman01}.  Because the starburst is both very strong and relatively nearby, it has been intensely studied at multiple wavelengths, including H$\alpha$, infrared, and radio emission, making it an ideal object to use in a comparison of SFRs from millimetre recombination line data to SFRs from data at other wavelengths.  Radio recombination line emission from NGC~5253 has been detected previously by \citet{mohan01} and \citet{rodriguezrico07} at lower frequencies, but they indicated that adjustments for masing and gas opacity would need to be taken into account to calculate SFR.  As millimetre recombination line emission is unaffected by these issues, it can provide more accurate measurements of the SFR.

We present here ALMA observations of the H30$\alpha$ line at a rest frequency of 231.90~GHz, with which we derive a SFR that directly traces the photoionizing stars while not being affected by dust attenuation.  We compare the ALMA-based SFR to SFRs from other wavebands to understand their effectiveness, and we also examine the efficacy of SFRs based on combining H$\alpha$ emission (a tracer of unobscured light from star forming regions) with infrared emission (a tracer of light absorbed and re-radiated by dust).  The ALMA observations and the SFRs derived from the line are presented in Section~\ref{s_alma}.  SFRs from optical, infrared, and radio emission as well as composite SFRs based on multiple wavebands are derived and compared to the H30$\alpha$ values in Section~\ref{s_othersfr}. Section~\ref{s_conclu} provides a summary of the results.

\section{ALMA data}
\label{s_alma}

\subsection{Observations and data processing}

Observations were performed with both the main (12~m) array and the Morita (7~m) Array with baselines ranging from 8 to 1568~m.  The observations with both arrays consist of two pointings centered on positions 13:39:55.9 -31:38:26 and 13:39:56.8 -31:38:32 in J2000 coordinates, which were used to map both emission from the compact central starburst and more diffuse, extended emission from gas to the southeast of the centre (see Miura et al. in preparation).  Additional data are also available from ALMA total power observations, but since the H30$\alpha$ emission is from a very compact source, the total power data will not substantially improve the detection of the line emission.  Moreover, inclusion of the total power data will complicate the data processing procedure, and these data currently cannot be used for continuum imaging.  Hence, we did not include the total power data in our analysis.

General information about each execution block used to construct the H30$\alpha$ image cube and continuum image is listed in Table~\ref{t_alma_obs}.  Information about the spectral set-up for each spectral window in each execution block is given in Table~\ref{t_alma_spw}.

\begin{table*}
\centering
\begin{minipage}{149mm}
\caption{ALMA observation information.}
\label{t_alma_obs}
\begin{tabular}{@{}lcccccccc@{}}
\hline
Array &
  Unique &
  Observing &
  On-source &
  Usable &
  uv &
  Bandpass &
  Flux &
  Phase \\
&
  identifier &
  dates &
  observing &
  antennas &
  coverage &
  calibrator &
  calibrator &
  calibrator \\
&
  &
  &
  time (min) &
  &
  (m) &
  &
  &
  \\
\hline
Morita &
  A002/X8440e0/X29c6 &
  15 Jun 2014 &
  30.7 &
  9 &
  8-48 &
  J1427-4206 &
  Ceres &
  J1342-2900 \\
Morita &
  A002/X9652ea/X5c3 &
  10 Dec 2014 &
  30.7 &
  9 &
  9-45 &
  J1337-1257 &
  Callisto &
  J1342-2900 \\
12~m &
  A002/X966cea/X25de &
  11 Dec 2014 &
  8.5 &
  37 &
  13-336 &
  J1337-1257 &
  Callisto &
  J1342-2900 \\
12~m &
  A002/Xa5df2c/X50ce &
  18 Jul 2015 &
  13.2 &
  39 &
  14-1512 &
  J1337-1257 &
  J1427-4206 &
  J1342-2900 \\
12~m &
  A002/Xa5df2c/X52fa &
  18 Jul 2015 &
  15.8 &
  39 &
  14-1568 &
  J1337-1257 &
  Titan &
  J1342-2900 \\
\hline
\end{tabular}
\end{minipage}
\end{table*}

\begin{table*}
\centering
\begin{minipage}{94mm}
\caption{ALMA spectral window settings.}
\label{t_alma_spw}
\begin{tabular}{@{}lcccc@{}}
\hline
Array &
  Unique &
  Frequency range$^a$ &
  Bandwidth &
  Number of \\
&
  identifier &
  (GHz) &
  (GHz) &
  channels \\
\hline
Morita &
  A002/X8440e0/X29c6 &
  229.216 - 231.208 &
  1.992 &
  2040 \\
&
  &
  230.577 - 232.569 &
  1.992 &
  2040 \\
&
  &
  243.594 - 245.586 &
  1.992 &
  2040 \\
Morita &
  A002/X9652ea/X5c3 &
  229.247 - 231.239 &
  1.992 &
  2040 \\
&
  &
  230.608 - 232.601 &
  1.992 &
  2040 \\
&
  &
  243.627 - 245.619 &
  1.992 &
  2040 \\
12~m &
  A002/X966cea/X25de &
  229.306 - 231.181 & 
  1.875 &
  1920 \\
&
  &
  230.667 - 232.542 & 
  1.875 &
  1920 \\
&
  &
  243.686 - 245.561 & 
  1.875 &
  1920 \\ 
12~m &
  A002/Xa5df2c/X50ce &
  229.270 - 231.145 &
  1.875 &
  1920 \\
&
  &
  230.630 - 232.505 &
  1.875 &
  1920 \\
&
  &
  243.647 - 245.522 &
  1.875 &
  1920 \\
12~m &
  A002/Xa5df2c/X52fa &
  221.269 - 231.144 &
  1.875 &
  1920 \\
&
  &
  230.630 - 232.505 &
  1.875 &
  1920 \\
&
  &
  243.647 - 245.522 &
  1.875 &
  1920 \\
\hline
\end{tabular}
$^a$ These values are the observed (sky) frequencies and not rest frequencies.
\end{minipage}
\end{table*}

The visibility data for each execution block was calibrated separately using the Common Astronomy Software Applications ({\sc CASA}) version 4.7.0.  To begin with, we applied amplitude corrections based on system temperature measurements to all data, and we applied phase corrections based on water vapour radiometer measurements to the 12~m data.  For the data with baselines greater than 1000~m, we also applied antenna position corrections.  Following this, we visually inspected the visibility data and flagged data with noisy or abnormal phase or amplitude values as well as atmospheric lines centered at 231.30~GHz.  Next, we derived and applied calibrations for the bandpass, phase, and amplitude.  The Butler-JPL-Horizons 2012 models were used to obtain the flux densities for Callisto and Titan.  J1427-4206 is one of the 43 quasars routinely monitored for flux calibration purposes as described within the ALMA Technical Handbook\footnote{https://almascience.eso.org/documents-and-tools/cycle4/alma-technical-handbook} \citep{magnum16}.  The {\sc getALMAFlux} task within the Analysis Utilities software package was used to calculate the flux density of J1427-4206 based on measurement in the ALMA Calibrator Source Catalogue\footnote{https://almascience.eso.org/alma-data/calibrator-catalogue}.  The version of the Butler-JPL-Horizons 2012 models implemented in {\sc CASA} 4.7.0 is known to produce inaccurate results for Ceres, so the data for the bandpass calibrator J1427-4206 and the estimated flux density from the {\sc getALMAFlux} task were used to flux calibrate the 15 June 2014 observation.  The uncertainty in band 6 flux calibration is specified in the ALMA Proposer's Guide\footnote{https://almascience.eso.org/documents-and-tools/cycle4/alma-proposers-guide} \citep{andreani16} as 10\%, but because the uncertainty in the estimated flux densities for J1427-4206 was 15\%, we used 15\% as the final calibration uncertainty.

The H30$\alpha$ image was created after subtracting the continuum from the visibility data in the spectral window containing the line.  The continuum was determined by fitting the visibility data at approximately 230.7-231.2~GHz and 231.7-232.5~GHz (in the Barycentric frame of reference) with a linear function; this frequency range avoids not only the H30$\alpha$ line but also the atmospheric absorption feature centered at 231.30~GHz.  After this, we concatenated the continuum-subtracted data  and then created a spectral cube using {\sc clean} within {\sc CASA}.  We used Briggs weighting with a robust parameter of 0.5, which is the standard weighting used when producing ALMA images of compact sources like the H30$\alpha$ source in NGC~5253.  After creating the image, we applied a primary beam correction.  The spectral channels in the cube have a width of 10~MHz (equivalent to a velocity width of 12.9 km s$^{-1}$) and range from sky frequencies of 231.40 to 231.80 GHz in the Barycentric frame.  The pixels have a size of 0.05~arcsec, and the imaged field was 100$\times$100~arcsec.  The reconstructed beam has a size of 0.21$\times$0.19~arcsec.

We also created a 231.6~GHz (observed frame) continuum image using the data from 230.7-231.2~GHz and 231.7-232.5~GHz.  We could have used the data from the spectral windows covering the CO and CS lines.  However, the potentially steep slope of the continuum at these frequencies (see the discussion in Section~\ref{s_images}), problems with observations on the shortest baselines in the spectral window covering the 243.6-245.6~GHz frequency range, and the relatively broad CO line emission made it difficult to create reliable continuum images using all spectral windows.  Because we found evidence for extended continuum emission, we used natural weighting when creating the final image.  The image dimensions are the same as for the H30$\alpha$ image cube.  The beam size is 0.28$\times$0.25~arcsec.

As a test of the flux calibration, we imaged J1427-4206 and J1337-1257 (another quasar monitored by ALMA for flux calibration purposes) using the same parameters that we used for creating the H30$\alpha$ image cube.  The measured flux densities differ by $<$10\% from the {\sc getALMAFlux} estimates, which is below our assumed flux calibration uncertainty of 15\%. 

We also checked the astrometry of each of the two longest-baseline observations by imaging the astrometry check source (J1339-2620) using the same parameters as for the continuum image, fitting the peak with a Gaussian function, and comparing the position to the coordinates in the ALMA Calibrator Source Catalogue.  The positions match to within one pixel, or 0.05~arcsec.  Equations in the ALMA Technical Handbook \citep{magnum16} yield a smaller value, but we will use 0.05~arcsec as the astrometric uncertainty.

\subsection{H30$\alpha$ and 231.6 GHz continuum images}
\label{s_images}

\begin{figure}
\epsfig{file=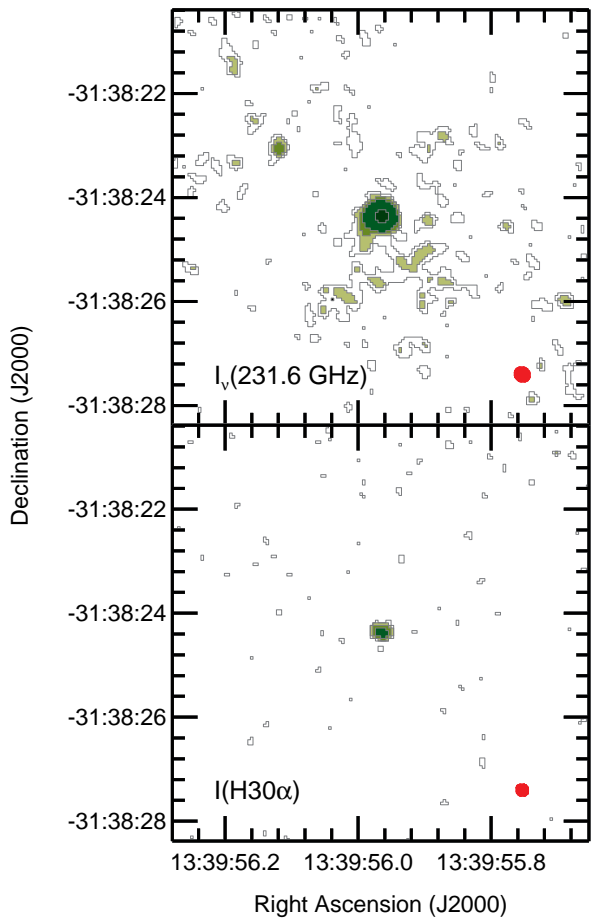,width=8.5cm}
\caption{Images of the central 8$\times$8~arcsec of NGC~5253 in 231.6~GHz (sky frequency) continuum emission (top) and H30$\alpha$ line emission (bottom).  The contours in each image show the 2$\sigma$, 3$\sigma$, 5$\sigma$, 10$\sigma$, and 100$\sigma$ detections levels in each image, where $\sigma$ is 1.0 Jy arcsec$^{-2}$ for the 231.6~GHz image and 0.69 Jy km s$^{1}$ arcsec$^{-2}$ for the H30$\alpha$ image.  The red ovals at the bottom right of each panel show the FWHM of the beam (0.28$\times$0.25~arcsec for the 231.6~GHz image and 0.21$\times$0.19~arcsec for the H30$\alpha$ image).}
\label{f_map}
\end{figure}

Figure~\ref{f_map} shows the 231.6~GHz (sky frequency) continuum and H30$\alpha$ spectral line images of the central 8$\times$8~arcsec of NGC~5253.  The H30$\alpha$ image is the integral of the continuum-subtracted flux between sky frequencies of 231.55 and 231.65~GHz.  Both sources show a very bright central peak at a right ascension of 13:39:55.965 and declination of -31:38:24.36 in J2000 coordinates.  No significant H30$\alpha$ emission is observed outside of the central peak.  The best-fitting Gaussian function to the H30$\alpha$ image has a FWHM 0.27$\times$0.21~arcsec, indicating that the H30$\alpha$ source may have a deconvolved angular size of $\sim$0.15~arcsec.  In the continuum image, a second unresolved source to the northeast of the centre is clearly detected at the 10$\sigma$ level.  Several other compact sources are detected in continuum at the 3-5$\sigma$ level, including a few sources outside the region shown in Figure~\ref{f_map}, and some diffuse emission near the 3$\sigma$ level is seen around the central peak, most notably immediately to south of the centre. 

\begin{figure}
\epsfig{file=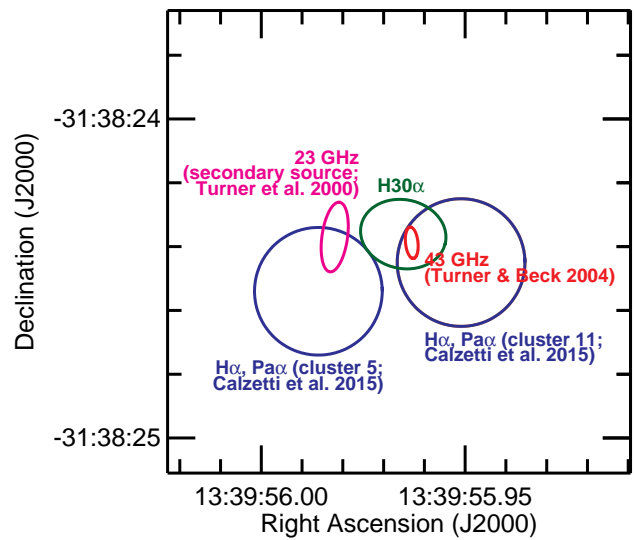,width=8.5cm}
\caption{ A map showing the positions of several sources detected in different bands in the centre of NGC~5253.  The H30$\alpha$ source is identified in green, and the ellipse represents the FWHM of a Gaussian function fit to the data, which is 0.27$\times$0.21~arcsec.  The blue circles identify the locations of the two brightest H$\alpha$ and Pa$\alpha$ sources imaged by \citet[][; blue]{calzetti15}, and the diameters of the circles represent the 0.20~arcsec astrometric uncertainty in the data.  The deconvolved size (0.099 $\times$ 0.039~arcsec) of the 43~GHz core imaged by \citet{turner04} is shown in red; this source also corresponds to the central source imaged at 15 and 23~GHz by \citet{turner00}.  A secondary 23~GHz source identified by \citet{turner04} is shown in magenta, and the ellipse diameter matches the 0.22 $\times$ 0.08~arcsec FWHM of the beam.}
\label{f_map_beam}
\end{figure}

Figure~\ref{f_map_beam} shows the location of the H30$\alpha$ source (which coincides with the location of the peak in 231.6~GHz emission) relative to other sources detected in observations with comparable angular resolutions.  The H30$\alpha$ source lies within 0.04~arcsec of the peak in 43~GHz emission measured by \citet{turner04}, which has the same position as the brightest 15 and 23~GHz sources detected in the high angular resolution radio observations presented by \citet{turner00}.  This offset is smaller than the astrometric uncertainty of 0.05~arcsec that we are using for the ALMA data.  \citet{turner00} also reported the detection of a second 23 GHz source at a location 0.21~arcsec east of the H30$\alpha$ source, although this source is not detected at 15~GHz, and it also is either a very weak detection or non-detection in the 43~GHz images shown by \citet{turner04}.  The difference between the peak fluxes of the primary and secondary 23~GHz sources is $\sim$6$\times$. If this secondary source has a corresponding H30$\alpha$ source, it would be difficult to separate the emission of this secondary source from the brighter H30$\alpha$ source given the small angular separation between the sources compared to the FWHM of the ALMA beam and the much lower flux expected from the secondary source relative to the brighter source.  Additional millimetre or radio observations with better angular resolutions and better sensitivities would be needed to confirm that the secondary source is present.

The two brightest H$\alpha$ and Pa$\alpha$ (1.876~$\mu$m) sources that were identified by \citet{calzetti15} using {\it Hubble} Space Telescope straddle the H30$\alpha$ source.  These sources, which were labelled clusters 5 and 11, have counterparts to sources identified by \citet{calzetti97}, \citet{alonsoherrero04}, \citet{harris04}, and \citet{degrijs13}.  \citet{calzetti15} had suggested that the brightest radio continuum sources actually corresponded to cluster 11 (as originally suggested by \citet{alonsoherrero04}), that the secondary 23~GHz source detected by \citet{turner00} at 23~GHz corresponds to cluster 5 and that the offset between the radio and optical/near-infrared data was related to systematic effects in the astrometry systems used to create the images.  It is therefore possible that the H30$\alpha$ source detected in the ALMA data corresponds to cluster 11.  However, clusters 5 and 11 are separated by a distance of 0.46 arcsec, so they should have been resolved into two separate components in the H30$\alpha$ map.  The Pa$\alpha$ fluxes measured by \citet{calzetti15} differ by only $\sim$25\%, but after they attempt to correct for dust absorption using models that account for the intermixing of the stars and dust,  the line fluxes are expected to differ by $\sim$6$\times$.  The peak of the H30$\alpha$ source is detected by ALMA at the 18$\sigma$ level, so a second photoionization region with at least one-sixth the flux of the main source would be detected at the 3$\sigma$ level.  

It is possible that the second source is even fainter than expected relative to the primary H30$\alpha$ source, which could occur if the primary source is more obscured than expected.  Alternately, it is possible that both sources lie at ends of a larger star forming complex that is heavily obscured in the optical and near-infrared bands, although the 0.46~arcsec separation between the two optical/near-infrared sources is larger than the $\sim$0.15~arcsec size of the H30$\alpha$ source we derived, which makes this second scenario less likely.

The coincidence of the H30$\alpha$ emission with the 231.6~GHz continuum emission provides additional support for the possibility that the central star forming region is very heavily obscured.  As discussed below, the 231.6~GHz continuum emission is expected to contain a combination of dust and free-free emission, although the relative contributions of each may be very uncertain.  If the 231.6~GHz emission is primarily from dust, then we would expect the optical and near-infrared light to be heavily obscured where most of the photoionizing stars are also located and for the Pa$\alpha$ and H$\alpha$ emission to be more easily detected at the fringes of the region.  In fact, \citet{calzetti15} noted that the Pa$\alpha$ emission is offset relative to the H$\alpha$ emission in their cluster 11, which would be consistent with this interpretation of the millimetre continuum and recombination line emission.  However, if the 231.6~GHz emission includes substantial free-free emission, then we would expect to easily detect a second photoionization region, even one that is 20$\times$ fainter than the central photoionizing region.

\begin{figure}
\epsfig{file=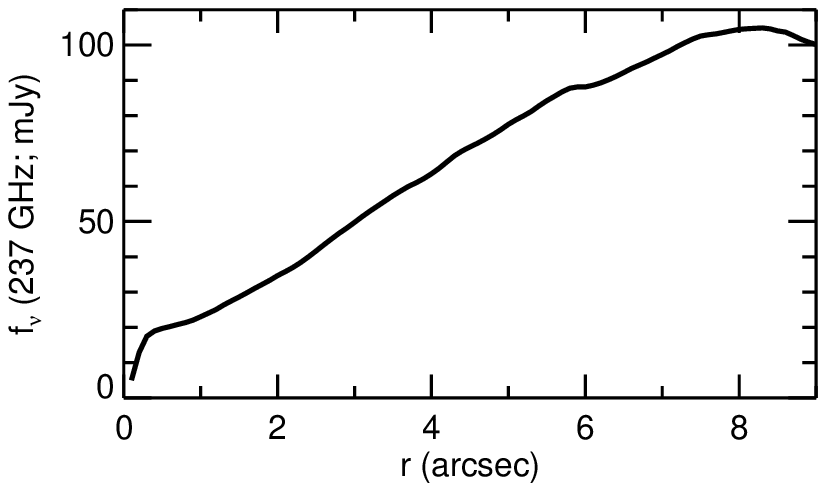}
\epsfig{file=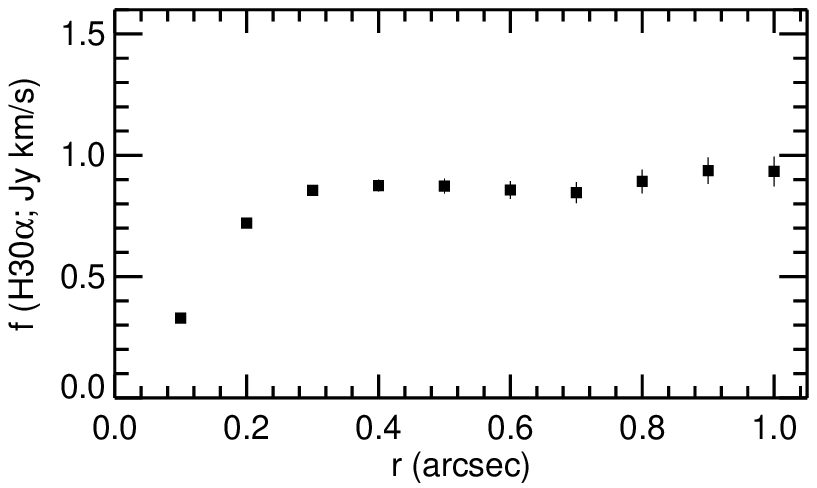}
\caption{Plots of the integrated 231.6~GHz continuum flux density (top) and integrated H30$\alpha$ flux (bottom) as a function of the radius of the measurement aperture.  The uncertainties in the continuum measurement related to random noise in the image are smaller than the width of the line.}
\label{f_curveofgrowth}
\end{figure}

To understand the distribution of the continuum and H30$\alpha$ emission, we measured integrated fluxes within apertures with radii varying from 0.1 arcsec, which is equivalent to the radius of the beam, to 9 arcsec, which is the radius at which we begin to measure artefacts related to the negative sidelobes of the central source.  These curve-of-growth profiles are shown in Figure~\ref{f_curveofgrowth}.  Most of the H30$\alpha$ flux from the unresolved central source falls within a radius of 0.3~arcsec, so we will use the measurements of the line flux from within that radius as the total flux.  The integrated continuum emission peaks at a radius of $\sim$8.5~arcsec.  Emission on angular scales larger than 17~arcsec is either resolved out, strongly affected by the negative sidelobes, or affected by the high noise at the edge of the primary beam.  While differences in the signal-to-noise ratio may explain why the continuum emission is detected at a larger radius than the H30$\alpha$ line, it is also possible that a significant fraction of the continuum emission may originate from dust within NGC~5253 that is distributed more broadly than the photoionized gas.

\begin{table*}
\centering
\begin{minipage}{109mm}
\caption{Continuum measurements of NGC~5253 at 230-250~GHz.}
\label{t_contcomp}
\begin{tabular}{@{}lccccc@{}}
\hline
Reference &
  Telescope &
  Frequency &
  Beam &
  Flux Density &
  Aperture \\
&
  &
  (GHz) &
  FWHM &
  (mJy) &
  Diameter\\
&
  &
  &
  (arcsec) &
  &
  (arcsec)\\
\hline
[this paper] &
  ALMA &
  231.6 &
  0.28$\times$0.25 &
  $104 \pm 16$ &
  17 \\
\citet{meier02} &
  OVRO &
  233.3 &
  $6.5 \times 4.5$ &
  $46 \pm 10$ &
  20 \\
\citet{vanzi04} &
  SEST &
  250 &
  11 &
  $114 \pm 4$ &
  11 \\
\citet{miura15} &
  SMA &
  230 & 
  $11 \times 4$ &
  $34 \pm 9$ &
  $11 \times 4$ \\
\hline
\end{tabular}
\end{minipage}
\end{table*}

The 231.6~GHz continuum flux density within a radius of 8.5~arcsec is 104$\pm$16~mJy.  A few other continuum flux densities at comparable frequencies have been published.  These measurements as well as details about the data are listed in Table~\ref{t_contcomp}.  

The ALMA measurement is very close to the Swedish-ESO Submillimetre Telescope (SEST) measurement from \citet{vanzi04}.  While the Vanzi et al. number is for both a smaller aperture and a higher frequency than the ALMA measurement, adjustments for both the measurement aperture and spectral slope should potentially cancel out.  

The \citet{meier02} flux density from the Owens Valley Radio Observatory (OVRO)and \citet{miura15} flux density from the Submillimeter Array (SMA) are both lower than the ALMA measurements by 2-3$\times$.  Both measurements have high uncertainties related to either calibration or detection issues, which could partly explain the mismatch between these data and ours.  The \citet{miura15} measurement is also for a smaller area; when we use a similar aperture to their beam size, we obtain a flux density of 47$\pm$7~mJy, which is near their measurement.  However, it is also likely that the OVRO and SMA data were insensitive to the faint, extended continuum emission from this source because of a combination of limited uv coverage and the broader beam size.  The extended emission observed by OVRO and SMA could have been smeared into the negative sidelobes of the central source or could have been redistributed on spatial scales larger than the largest angular scales measurable by the arrays.  The ALMA observations, which used both short and long baselines, can recover structures on the same angular scales as OVRO and SMA while not spreading the emission onto scales where it is not recoverable by the interferometer, and since the ALMA data have a smaller beam, any negative sidelobes will not cover a significant fraction of the diffuse, extended emission within a radius of 8.5~arcsec of the centre.

We therefore think the ALMA flux density, which is consistent with the SEST measurement, should be fairly reliable.  Having said this, we will emphasize that much of the continuum emission in our image has a surface brightness at $<$5$\sigma$; more sensitive measurements would be needed to confirm our results.  

The emission at 231.6~GHz could originate from a variety of sources.  The amount of dust emission at 231.6~GHz can be estimated by extrapolating the modified blackbody function from \citet{remyruyer13} that was fit to the 100-500~$\mu$m {\it Herschel} Space Observatory data.  This gives a flux density of 63~mJy, but the uncertainties in the parameters for the best fitting function indicate that the uncertainty is$\sim$2$\times$.  Free-free emission as well as a few more exotic emission mechanisms could contribute to the emission at 231.6~GHz.  While a more in-depth analysis of the SED is beyond the scope of this paper, we discuss this topic further in Section~\ref{s_discuss_mmradio}.

\begin{figure}
\epsfig{file=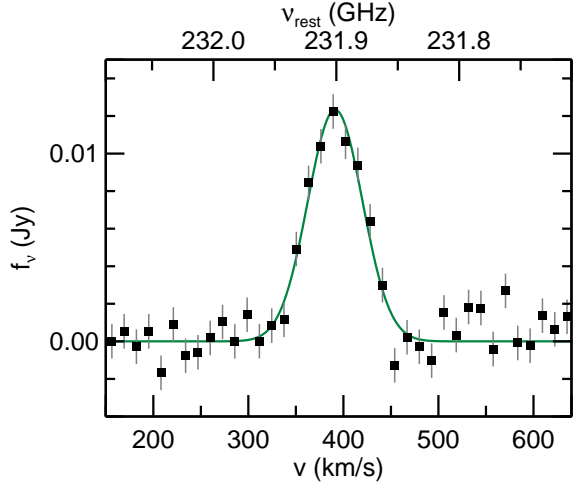}
\caption{The continuum-subtracted spectrum of the centre of NGC~5253 showing the H30$\alpha$ line emission.  This spectrum was measured within a region with a radius of 0.5~arcsec.  The green line shows the best-fitting Gaussian function, which has a mean relativistic velocity in the Barycentric frame of 391$\pm$2~km~s$^{-1}$ and a FWHM of 68$\pm$3~km~s$^{-1}$.}
\label{f_spec}
\end{figure}

Figure~\ref{f_spec} shows the portion of the continuum-subtracted spectrum that includes the H30$\alpha$ line emission.  The line has a mean relativistic velocity in the Barycentric reference frame of 391$\pm$2~km~s$^{-1}$ and a FWHM of 68$\pm$3~km~s$^{-1}$.  The integral of the line is 0.86~Jy~km~s$^{-1}$ with a measurement uncertainty of 0.04~Jy~km~s$^{-1}$ and a calibration uncertainty of 0.13~Jy (15\%).  The spectral window covering the H30$\alpha$ line does not include any other detectable spectral lines.  Miura et al. (in preparation) will discuss the CO (2-1) line emission and any other spectral lines detected in the other spectral window.

\subsection{SFR from the H30$\alpha$ data}
\label{s_alma_sfr}

For this analysis, we assume that the nuclear starburst detected in H30$\alpha$ data contains most of the photoionizing stars in the galaxy and that a SFR derived from it will be representative of the global SFR.  While the source is near the brightest clusters found in H$\alpha$ and Pa$\alpha$ emission, fainter $<$5~Myr clusters and diffuse emission are found outside the central starburst \citep{calzetti04, calzetti15, harris04}.  Although most of this emission should fall within the region imaged by ALMA, the lack of H30$\alpha$ emission detected from these fainter sources could cause the SFR from the H30$\alpha$ emission to be biased downwards.  Given the central concentration of the H$\alpha$ and Pa$\alpha$ emission, however, the bias may not be too severe to significantly affect comparisons to other star formation tracers.

The H30$\alpha$ flux can be converted to a photoionizing photon production rate $Q$ using
\begin{equation}
\begin{split}
 \frac{\mbox{Q}(\mbox{H30}\alpha)}{\mbox{s}^{-1}}=
  3.99\times10^{30}
  \left[\frac{\alpha_B}{\mbox{ cm}^3\mbox{ s}^{-1}}\right]
  \ \ \ \ \ \ \ \ \ \ \ \ \ \ \ \ \ \ \ \ \ \ \ \ \ \ \ \ \ \ \\
  \times
  \left[\frac{\epsilon_\nu}{\mbox{erg s}^{-1}\mbox{ cm}^{-3}}\right]^{-1}
  \left[\frac{\nu}{\mbox{~GHz}}\right]
  \left[\frac{D}{\mbox{ Mpc}}\right]^{2}
  \left[\frac{\int f_\nu(\mbox{line}) dv}{\mbox{Jy km s}^{-1}}\right]
\end{split}
\label{e_q_h30a}
\end{equation}
based on equations from \citet{scoville13}.  The effective recombination coefficient ($\alpha_B$) and emissivity ($\epsilon_\nu$) terms in this equation depend on the electron density and temperature.  Both term varies by less than 15\% for electron densities between $10^2$ and $10^5$cm$^{-3}$.  The terms do vary significantly with electron temperature between 3000 and 15000~K, and the resulting $Q$ can change by a factor of $\sim$2.4 depending on which temperature is selected.  In analyses at lower frequencies, where the continuum emission is dominated by free-free emission, it is possible to use the line-to-continuum ratio to estimate electron temperatures.  However, thermal dust emission comprises a significant yet poorly-constrained fraction of the 231.6~GHz continuum emission in the central starburst, so this method would produce questionable results.  Instead, we use 11500~K for the electron temperature, which is based on [O{\small III}] measurements near the centres of the brightest optical recombination line sources within NGC~5253 from \citet{kobulnicky97}, \citet{lopezsanchez07}, \citet{guseva11}, and \citet{monrealibero12}.  For the electron density, we use 600 cm$^{-3}$, which is the mean of electron densities based on measurements of the [O{\small II}] and [S{\small II}] from these same regions as given by \citet{lopezsanchez07}, \citet{guseva11}, and \citet{monrealibero12}.  Although the electron density measurements from these three studies vary by a factor of $\sim$3, probably because of issues related to the position and size of the measurement apertures as discussed by \citet{monrealibero12}, the resulting $Q$ values should be unaffected by the relatively large disagreement in these measurements.  Dust extinction could have affected both the electron temperature and density estimates, but we do not expect extinction affects to alter the data to a degree where our calculations are affected.  We interpolated among the $\alpha_B$ and $\epsilon_\nu$ terms published by \citet{storey95} to calculate specific values for these terms based on our chosen electron temperature and density.  Based on these terms and our H30$\alpha$ flux, we calculated $Q$ to be (1.9$\pm$0.3)$\times$$10^{52}$~s$^{-1}$. 

The $Q$ value can be converted to SFR (in M$_\odot$ yr$^{-1}$) using a simple scaling term, but this term depends upon the characteristics of the stellar population within the star forming regions.  Using {\sc Starburst99} \citep{leitherer99}, \citet{murphy11} derived a scaling term of $7.29\times10^{-54}$ for solar metallicity (defined in older versions of {\sc Starburst99} as $Z$=0.020) and a \citet{kroupa02} initial mass function (IMF)\footnote{The Kroupa IMF used in {\sc Starburst99} is defined as having an index of 2.3 between 100~M$_\odot$ (the IMF upper mass boundary) and 0.5~M$_\odot$ and an index of 1.3 between 0.5~M$_\odot$ and 0.1~M$_\odot$ (the IMF lower mass boundary).}.  However, the conversion factor is dependent on metallicity.  \citet{monrealibero12} reported that 12+log(O/H) for NGC~5253 is 8.26$\pm$0.04 (on a scale where solar metallicity corresponds to 12+log(O/H)=8.66 and $Z$=0.014), which is consistent with older results \citep{kobulnicky97, lopezsanchez07, guseva11}.  Assuming that all other abundances scale with the O/H ratio, this oxygen abundance is equivalent to $Z$$\cong$0.0056.

The current version (7.0.1) of {\sc Starburst99} \citep{leitherer14} provides the Geneva 2012/13 evolutionary tracks, which include stellar rotation \citep{ekstrom12, georgy12}, as an option for the simulations.  Stellar rotation enhances convection and also metallicity at the surfaces of the stars, which makes stars hotter and therefore affects the conversion between $Q$ and SFR.  Currently, {\sc Starburst99} includes tracks for no rotation and rotation at 40\% of the break up velocity.  These two extremes are expected to bracket the actual rotation velocities of typical stellar populations.  The current version of {\sc Starburst99} also only includes Geneva tracks with rotation for $Z=0.002$ and $Z=0.014$.  We assume that the logarithm of $Q$ scales linearly with $Z$ when deriving conversion factors.  Based on {\sc Starburst99} results using the 1994 versions of the Geneva tracks \citep{shaller92, charbonnel93, schaerer93a, schaerer93b}, this interpolation should be accurate to within 5\%.

\begin{figure*}
\epsfig{file=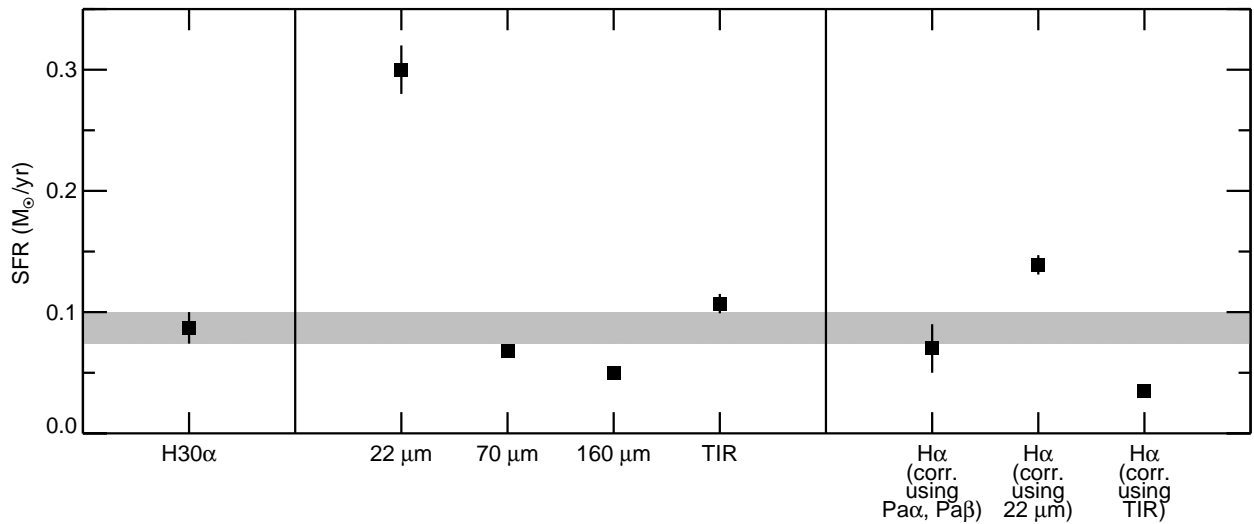}
\caption{A graphical depiction of the SFRs for NGC~5253.  These are values calculated using corrections to account for the metallicity of NGC~5253 ($Z=0.0056$) and stellar rotation effects; they correspond to the values in the rightmost column in Table~\ref{t_sfr}.  Uncertainties for the individual data points (when they are larger than the symbols in this plot) correspond to the measurement uncertainties and uncertainties related to the correction factors in Table~\ref{t_sfrcorrfac} but do not include uncertainties in the conversion factors between the fluxes and SFR.  The grey band corresponds to the mean and 1$\sigma$ uncertainties in the SFR from the H30$\alpha$ data.}
\label{f_sfr}
\end{figure*}

We report three versions of the SFR: a version based on solar metallicity and no rotation using the \citet{murphy11} conversion; a version based on the Geneva 2012/13 tracks for $Z$=0.0056 with no stellar rotation; and a version based on the Geneva 2012/13 tracks for $Z$=0.0056 where we use the average of model results for no rotation and 40\% of the break up velocity.  The last two numbers are derived from {\sc Starburst99} simulations using a Kroupa IMF and should be applicable to scenarios where star formation has been continuous for $>$10~Myr.  The three conversion factors are listed in Table~\ref{t_qsfrconversion}, with more general-purpose correction factors for metallicity and rotation effects listed in Table~\ref{t_sfrcorrfac}.  The SFRs based on Equation~\ref{e_q_h30a} and the conversion factors between $Q$ and SFR are reported in Table~\ref{t_sfr}.  Additionally, Figure~\ref{f_sfr} shows a graphical comparison between the SFR from the H30$\alpha$ data and the SFRs calculated from other star formation tracers as described in the following section.  In accounting for metallicity effects (using the new Geneva tracks), the resulting SFR decreases by $\sim$25\%.  When rotation is incorporated, the SFR decreases by an additional $\sim$10\% relative to the solar value without rotation.

\begin{table}
\caption{Conversions between $Q$ and SFR.}
\label{t_qsfrconversion}
\begin{center}
\begin{tabular}{@{}lc@{}}
\hline
Model &
  Conversion from $Q$ to SFR \\
&
  ( M$_\odot$ yr$^{-1}$ / (s$^{-1}$) )\\
\hline
Solar metallicity, no stellar rotation &
  $7.29\times10^{-54~a}$\\
$Z$=0.0056$^b$, no stellar rotation &
  $5.40\times10^{-54}$\\
$Z$=0.0056$^b$, with stellar rotation &
  $4.62\times10^{-54}$\\
\hline
\end{tabular}
\end{center}
$^a$ This conversion is from \citet{murphy11}.\\
$^b$ These metallicities are based on a scale where solar metallicity is 0.014.
\end{table}

\begin{table*}
\centering
\begin{minipage}{123mm}
\caption{Corrections in SFR relative to solar metallicity scenario with no stellar rotation.}
\label{t_sfrcorrfac}
\begin{tabular}{@{}cccc@{}}
\hline
General &
  Specific &
  \multicolumn{2}{c}{Correction factor$^a$} \\
waveband &
  star formation &
  $Z$=0.0056 &
  $Z$=0.0056 \\
&
  metrics &
  no stellar rotation$^b$ &
  with stellar rotation$^b$ \\
\hline
Recombination lines &
  H$\alpha$, H30$\alpha$ &
  0.74 &
  0.63 \\
Infrared &
  22~$\mu$m, 70~$\mu$m, 160~$\mu$m, total infrared &
  $0.90 \pm 0.03$ &
  $0.72 \pm 0.03$ \\
\hline
\end{tabular}
$^a$ Derived SFRs should be multiplied by these correction factors.\\
$^b$ On this scale, solar metallicity corresponds to $Z$=0.014, and $Z$=0.0056 is the approximate metallicity of NGC~5253 based on 12+log(O/H) measurements.
\end{minipage}
\end{table*}

\begin{table*}
\centering
\begin{minipage}{119mm}
\caption{SFRs calculated for NGC 5253.}
\label{t_sfr}
\begin{tabular}{@{}lccc@{}}
\hline
Waveband &
  \multicolumn{3}{c}{SFR (M$_\odot$~yr$^{-1}$)$^a$}\\
&
  Solar metallicity &
  $Z$=0.0056 &
  $Z$=0.0056 \\
&
  no stellar rotation &
  no stellar rotation &
  with stellar rotation \\
\hline
H30$\alpha$ &
  $0.14 \pm 0.02$ &
  $0.102 \pm 0.015$ &
  $0.087 \pm 0.013$ \\
\hline
22~$\mu$m &
  $0.41 \pm 0.02$ &
  $0.37 \pm 0.02$ &
  $0.30 \pm 0.02$ \\
70~$\mu$m &
  $0.095 \pm 0.004$ &
  $0.086 \pm 0.005$ &
  $0.068 \pm 0.004$ \\
160~$\mu$m &
  $0.070 \pm 0.003$ &
  $0.063 \pm 0.003$ &
  $0.050 \pm 0.003$ \\
Total infrared &
  $0.148 \pm 0.009$ &
  $0.133 \pm 0.009$ &
  $0.107 \pm 0.008$ \\
\hline
H$\alpha$ (corrected using Pa$\alpha$, Pa$\beta$ data)$^b$ &
  $0.11 \pm 0.03$ &
  $0.08 \pm 0.02$ &
  $0.07 \pm 0.02$ \\
H$\alpha$ (corrected using 22~$\mu$m data)&
  $0.221 \pm 0.012$ &
  $0.164 \pm 0.009$ &
  $0.139 \pm 0.008$ \\
H$\alpha$ (corrected using total infrared data) &
  $0.055 \pm 0.04$ &
  $0.041 \pm 0.003$ &
  $0.035 \pm 0.003$ \\
\hline
\end{tabular}
$^a$ All uncertainties incorporate measurement uncertainties and uncertainties related to the correction factors in Table~\ref{t_sfrcorrfac}.  The uncertainties do not include the uncertainties in the multiplicative factos applied when converting from the fluxes in the individual bands to SFR.\\
$^b$ The extinction corrections for these data were calculate and applied by \citet{calzetti15}.\\
\end{minipage}
\end{table*}

\section{Other star formation tracers}
\label{s_othersfr}

Having derived the SFR from the H30$\alpha$ data, we calculated SFRs using other publicly-available infrared and H$\alpha$ data.  While the H30$\alpha$ emission primarily originates from stars younger than 5~Myr, other star formation tracers may trace stellar populations with different ages, so the derived SFRs could differ if the rate has changed significantly over time.  SFRs based on most star formation tracers are also affected by IMF, metallicity, and stellar rotation effects.  Most current conversions, including all of the ones we used, are calibrated for a Kroupa IMF.  However, the SFR equations are usually calibrated for solar metallicities and do not account for stellar rotation.  We therefore report SFRs both using the original equations for solar metallicity and using conversions modified for the metallicity of NGC~5253.  For the lower metallicity scenario, we include SFRs with and without incorporating stellar rotation effects.  All corrections are derived using version 7.0.1 of the {\sc Starburst99} models and are applicable to scenarios with a Kroupa IMF and continuous star formation older than 10~Myr.  The corrections for stellar rotation are the average of values for no rotation and 40\% of the break-up velocity.

\subsection{Comparisons of H30$\alpha$ results to other results from millimetre and radio data}
\label{s_discuss_mmradio}

The H30$\alpha$ results can be compared to some of the published $Q$ based on other radio data analyses.  To start with, two papers have published analyses based on radio recombination lines.  \citet{mohan01} published multiple values of $Q$ based on different models applied to H92$\alpha$ (8.31 GHz) data; these $Q$ values range from 0.9$\times$$10^{52}$ to 2$\times$$10^{52}$~s$^{-1}$ when rescaled for a distance of 3.15~Mpc.  \citet{rodriguezrico07} using H53$\alpha$ (42.95~GHz) data obtained a $Q$ of 1.2$\times$$10^{52}$~s$^{-1}$.  Most of these measurements are lower than the value of (1.9$\pm$0.3)$\times$$10^{52}$~s$^{-1}$ from the H30$\alpha$ data, although Model IV from \citet{mohan01} produced a slightly higher $Q$, and most other results differ by less than 2$\times$.  Both the \citet{mohan01} and \citet{rodriguezrico07} $Q$ values are based on models that have attempted to account for masing and gas opacity effects, but it is possible that they were not always able to correct for these effects accurately.  Nonetheless, some fine-tuning of the models of the lower frequency recombination line emission may yield more accurate $Q$ and SFR values from the higher order recombination lines.

A series of analyses of the 23 - 231 GHz continuum emission published by \citet{turner00}, \citet{meier02}, \citet{turner04}, and \citet{miura15} treated the emission at these frequencies as originating mainly from free-free emission.  Based on this assumption, these observations gave $Q$ values ranging from 2.4$\times$$10^{52}$ to 4.8$\times$$10^{52}$ s$^{-1}$ after being rescaled for a distance of 3.15~Mpc.  Some of these values are significantly higher than the value of (1.9$\pm$0.3)$\times$$10^{52}$~s$^{-1}$ from the H30$\alpha$ line emission.  As discussed in Section~\ref{s_images}, thermal dust emission could produce more than half of the observed emission at 231~GHz, and this probably resulted in the high $Q$ values calculated by \citet{meier02} and \citet{miura15} based on photometry as similar frequencies.  The highest $Q$ from \citet{turner04} is based on 43~GHz emission measured within a region with a diameter of 1.2~arcsec.  \citet{rodriguezrico07} determined that only $\sim$20\% of the 43~GHz emission originated from a region with a diameter of 0.4~arcsec, comparable to the region where we detected H30$\alpha$ emission.  If we adjust the value of $Q$ from \citet{turner04} to account for the aperture effects, the value drops to $\sim$1.0$\times$$10^{52}$ s$^{-1}$, which is nearly 2$\times$ lower than what we measure from the H30$\alpha$ source.

\citet{meier02} and \citet{rodriguezrico07} both indicate that optical thickness effects could alter the free-free emission, which complicates the conversion from continuum emission to $Q$.  Additionally, it is possible that other emission sources that are not understood as well from a physical standpoint, such as the "submillimetre excess" emission that has been identified in many other low-metallicity dwarf galaxies \citep[e.g. ][]{galametz11,remyruyer13} or anomalous microwave emission \citep[e.g. ][]{murphy10, dickinson13}, could contribute to the emission in the millimetre bands.  The data from the analysis in our paper are insufficient for determining how any of these phenomena affect the radio or millimetre emission from NGC~5253 or affect the calculation of $Q$ from radio or millimetre data.  A deeper analysis of the SED, potentially including the reprocessing and analysis of ALMA 86 - 345~GHz data acquired in 2014 and later as well as a re-examination of archival radio data from the Very Large Array, would be needed to identify the various emission components in the 1-350~GHz regime as well as to determine how to convert the continuum emission to SFR.  However, this is beyond the scope of our paper.

Aside from the calculations described above, a widely-used conversion between radio continuum emission and SFR relies upon the empirical correlation between 1.4~GHz emission, which is an easily-observable radio continuum band, and far-infrared emission.  We can only find published 1.4~GHz flux densities integrated over an area much broader than the aperture used for the H30$\alpha$ measurement, and the radio emission is very extended compared to the central starburst \citep[e.g. ][]{turner98, rodriguezrico07}.  Hence, any SFR calculated using the available globally-integrated flux density will also be affected by aperture effects, so it would be inappropriate to list it alongside the other SFRs in Table~\ref{t_sfr}.  Having said this, if we use 
\begin{equation}
\frac{\mbox{SFR}(\mbox{1.4 GHz})}{\mbox{M}_\odot\mbox{ yr}^{-1}}
  = 0.0760 \left[\frac{D}{\mbox{Mpc}}\right]^{2}
  \left[\frac{f_\nu(1.4 \mbox{ GHz})}{\mbox{Jy}}\right],
\label{e_sfr_1.4ghz}
\end{equation}
which is based on the conversion from \citet{murphy11}, and the globally-integrated measurement of $84.7 \pm 3.4$~mJy from the 1.4~GHz NRAO VLA Sky Survey \citep{condon98}, we obtain a SFR of $0.064 \pm 0.003$~M$_\odot$~yr$^{-1}$.

The SFR from the 1.4~GHz emission is lower than the SFR from the H30$\alpha$ data even though the 1.4~GHz emission is measured over a much larger area.  The 1.4~GHz band has been calibrated as a star formation tracer using data for solar metallicity galaxies and relies upon a priori assumptions about the relative contributions of free-free and synchrotron emission to the SED.  The ratio of these forms of emission can vary between spiral and low mass galaxies, but the relation between infrared and radio emission has been found to staty linear in low metallicity galaxies in general \citep[e.g. ][ and references therein]{bell03,wu07}.  In NGC~5253 specifically, the relative contribution of synchrotron emission to the 1.4~GHz band is very low \citep{rodriguezrico07}, which is probably sufficient to cause the SFR from the globally-integrated 1.4~GHz measurement to fall below the SFR from the nuclear H30$\alpha$ measurement as well as the SFR from the total infrared flux (see Section~\ref{s_othersfr_ir}).

In any case, it appears that Equation~\ref{e_sfr_1.4ghz} simply is not suitable for NGC~5253, as it evidently deviates from the empirical relations between radio emission and either infrared flux or SFR.  An SFR based on the spectral decomposition of the SED and the analysis of its components would potentially yield more accurate results, but as we stated above, this analysis is beyond the scope of this paper.

\subsection{Comparisons of H30$\alpha$ and infrared dust continuum results}
\label{s_othersfr_ir}

\subsubsection{SFR calculations}
\label{s_othersfr_ircalc}

For calculating SFRs based on infrared flux densities, we used globally-integrated values, mainly because the infrared emission appears point-like in most data and because no data exist with angular resolutions comparable to the ALMA data.  Figures~\ref{f_map} and \ref{f_curveofgrowth} show that a significant fraction of the cold dust emission originates from an extended region outside the central starburst.  However, the 8~$\mu$m image from \citet{dale09} shows that most of the emission in that band (a combination of hot dust and polycyclic aromatic hydrocarbon (PAH) emission as well as a small amount of stellar emission) originates from a central unresolved source with a diameter smaller than 2~arcsec.  The emission at mid-infrared wavelengths should be very compact and therefore should be directly related to the luminosity of the central starburst seen in H30$\alpha$ emission.  The SFR from the total infrared emission could be affected by diffuse dust heated by older stellar populations outside the centre and therefore could be higher.  While we do not apply any corrections for this diffuse dust, we discuss the implications of this more in Section~\ref{s_othersfr_irdiscuss}.

\begin{table}
\caption{Infrared data for NGC 5253.}
\label{t_data_ir}
\begin{center}
\begin{tabular}{@{}lc@{}}
\hline
Wavelength &
  Flux \\
($\mu$m) &
  density$^a$ (Jy)\\
\hline
12 &
  $1.86 \pm 0.08$ \\
22 &
  $12.4\pm0.7$ \\
70 &
  $33.0 \pm 1.6$ \\
100 &
  $33.2 \pm 1.6$ \\
160 &
  $22.4 \pm 1.2$ \\
250 &
  $7.3 \pm 0.6$ \\
350 &
  $3.3 \pm 0.3$ \\
500 &
  $1.04 \pm 0.09$\\
\hline
\end{tabular}
\end{center}
\end{table}

We calculated SFRs based on flux densities measured in individual bands as well as total infrared fluxes based on integrating the SED between 12 and 500~$\mu$m, which covers most of the dust continuum emission.  Table~\ref{t_data_ir} shows the data we used in this analysis.

For the 12 and 22~$\mu$m bands, we made measurements within Wide-field Infrared Survey Explorer \citep[WISE; ][]{wright10} images from the AllWISE data release.  Although 24~$\mu$m flux densities based on {\it Spitzer} data have been published by \citet{engelbracht08} and \citet{dale09}, the centre of NGC~5253 is saturated in the 24~$\mu$m image \citep{bendo12}.  Hence, we used the WISE 22~$\mu$m data instead and assumed that the conversions from WISE 22~$\mu$m flux densities to SFR will be the same as for {\it Spitzer} 24~$\mu$m data.  The 12 and 22~$\mu$m flux densities were measured within circles with radii of 150 arcsec; this is large enough to encompass the optical disc of the galaxy as well as the beam from the central source, which has a FWHM of 12~arcsec at 22~$\mu$m \citep{wright10}.  The backgrounds were measured within annuli with radii of 450 and 500~arcsec and subtracted before measuring the flux densities.  Calibration uncertainties are 5\% at 12~$\mu$m and 6\% at 22~$\mu$m.  Colour corrections from \citet{wright10}, which change the flux densities by $\ltsim$10\%, are applied based on spectral slopes proportional to $\nu^{-3}$ at 12~$\mu$m and $\nu^{-2}$ at 22~$\mu$m, which is based on the spectral slopes between 12 and 22 and between 22 and 70~$\mu$m.

For the 70, 100, 160, 250, 350, and 500~$\mu$m measurements, we used the globally-integrated flux densities measured from {\it Herschel} data by \citet{remyruyer13}.  These data include no colour corrections, so we applied corrections equivalent to that for a point-like modified blackbody with a temperature of 30~K and an emissivity that scales as $\nu^2$, which is equivalent to the shape of the modified blackbody fit by \citet{remyruyer13} to the data.  The 70, 100, and 160~$\mu$m colour corrections, which change the flux densities by $\ltsim$5\%, are taken from \citet{muller11}\footnote{http://herschel.esac.esa.int/twiki/pub/Public/PacsCalibrationWeb\\ /cc\_report\_v1.pdf}, and the 250, 350, and 500~$\mu$m colour corrections, which change the flux densities by $\sim$10\%, are taken from \citet{valtchanov17}\footnote{http://herschel.esac.esa.int/Docs/SPIRE/spire\_handbook.pdf}.  These colour-corrected data are listed in Table~\ref{t_data_ir}.

While multiple methods have been derived for calculating the total infrared flux using a weighted sum of flux densities measured in multiple other individual bands \citep[e.g. ][]{dale02,boquien10,galametz13,dale14}, most of these derivations are calibrated for galaxies with SEDs similar to those of spiral galaxies.  The dust in NGC~5253 is much hotter than in than in most spiral galaxies, and in particular, the 22/70~$\mu$m ratio of 0.38 in NGC~5253 is higher than the 24/70~$\mu$m ratio of 0.05-0.10 found in many spiral galaxies \citep[e.g. ][]{dale07,bendo12}.  We therefore calculated the total infrared flux by first linearly interpolating between the logarithms of the monochromatic flux densities as a function of the logarithm of the wavelength and then integrating under the curve.  This gave a total infrared flux of (3.2$\pm$0.2)$\times$$10^{14}$~Jy~Hz, with the uncertainties derived using a monte carlo analysis.

Most conversions from infrared emission to SFR work given the assumption that star forming regions produce most of the observed bolometric luminosity of galaxies and that dust absorbs and re-radiates virtually all of the emission from the star forming regions.  Additionally, the conversions of measurements from individual infrared wavebands to SFR is based on the assumption that the SED does not change shape and that the flux densities in the individual band scale linearly with the total infrared flux.  For this analysis, we used 
\begin{equation}
  \frac{\mbox{SFR}(22\mu\mbox{m})}{\mbox{M}_\odot\mbox{ yr}^{-1}}=
  2.44\times10^{-16} \left[\frac{D}{\mbox{Mpc}}\right]^{2}
  \left[\frac{\nu f_\nu(22\mu\mbox{m})}{\mbox{Jy Hz}}\right]
\label{e_sfr_22}
\end{equation}
from \citet{rieke09},
\begin{equation}
  \frac{\mbox{SFR}(70\mu\mbox{m})}{\mbox{M}_\odot\mbox{ yr}^{-1}}=
  7.04\times10^{-17} \left[\frac{D}{\mbox{Mpc}}\right]^{2}
  \left[\frac{\nu f_\nu(70\mu\mbox{m})}{\mbox{Jy Hz}}\right]
\label{e_sfr_70}
\end{equation}
from \citet{calzetti10},
\begin{equation}
  \frac{\mbox{SFR}(160\mu\mbox{m})}{\mbox{M}_\odot\mbox{ yr}^{-1}}=
  1.71\times10^{-16} \left[\frac{D}{\mbox{Mpc}}\right]^{2}
  \left[\frac{\nu f_\nu(160\mu\mbox{m})}{\mbox{Jy Hz}}\right]
\label{e_sfr_160}
\end{equation}
from \citet{calzetti10}, and
\begin{equation}
\begin{split}
  \frac{\mbox{SFR}(\mbox{total infrared})}{\mbox{M}_\odot\mbox{ yr}^{-1}}
  \ \ \ \ \ \ \ \ \ \ \ \ \ \ \ \ \ \ \ \ \ \ \ \ \ \ \ \ \ \ \ \ \ \ \ \ \ \ \ \ \ \ \ \ \ \ \ \ \ \ \ \ \ \ \ \ \ \ \ \ \ \ \ \\
  =4.66\times10^{-17}
  \left[\frac{D}{\mbox{Mpc}}\right]^{2}
  \left[\frac{f(\mbox{total infrared})}{\mbox{Jy Hz}}\right]
\end{split}
\label{e_sfr_tir}
\end{equation}
from \citet{kennicutt12} based on derivations from \citet{hao11} and \citet{murphy11}.  The SFRs calculation using these equations are listed in Table~\ref{t_sfr}.

While the conversion for total infrared flux is derived using {\sc Starburst99} for solar metallicity and older (but unspecified) stellar evolutionary tracks, the conversions for the 22, 70, and 160~$\mu$m bands are based on part upon empirical relations between the emission in those individual bands and other star formation tracers.  However, all of these conversions are based upon the assumption that the infrared flux as measured in an individual band or as integrated over a broad wavelength range will scale with the bolometric luminosity.  Based on both the Geneva 1994 and 2012/13 tracks implemented in {\sc Starburst99} version 7.0.1, the conversions factors in Equations~\ref{e_sfr_22}-\ref{e_sfr_tir} should be multiplied by 0.90 to correct for the lower metallicity in NGC~5253.  To account for metallicity and rotation in the same way as we did for the recombination lines, the conversions need to be multiplied by 0.72.  Versions of the SFRs with these corrections applied are listed in Table~\ref{t_sfr} alongside the value calculated assuming solar metallicity and no stellar rotation.

\subsubsection{Discussion}
\label{s_othersfr_irdiscuss}

The 22~$\mu$m flux density yielded a SFR that was $\sim$3$\times$ higher than the H30$\alpha$ SFR and is also significantly higher than the SFRs calculated using most other methods.  The aberrant SFR from the mid-infrared data is a consequence of the low metallicity of NGC~5253.  As stated in the previous section, the conversions from flux densities in individual infrared bands to SFRs are based upon the key assumptions that the total infrared flux originates from light absorbed from star forming regions and that the individual bands will scale linearly with the total infrared flux.  When the second condition is not met, the SFRs from individual infrared bands will be inaccurate.  This problem had been anticipated for low metallicity galaxies like NGC~5253 \citep[e.g. ][]{calzetti10}.  Low metallicity galaxies contain less interstellar dust, so the light from star forming regions is not attenuated as much as it is in larger galaxies.  As a result, the dust that is present is irradiated by a relatively hard and strong radiation field, which makes the dust warmer than in spiral galaxies \citep{hunt05, engelbracht08, rosenberg08, hirashita09, remyruyer13}.  The resulting change in the shape of the infrared SED results in biasing the SFR from 22~$\mu$m data upwards.    

The 70~$\mu$m flux density yielded a SFR that was 1-2$\sigma$ smaller than the H30$\alpha$ value (depending on which versions from in Table~\ref{t_sfr} are compared), while the SFR from the 160~$\mu$m data was approximately half of the H30$\alpha$ value.  This change in the calculated SFR with increasing wavelength is clearly a consequence of the unusually hot dust within NGC~5253.  The longer wavelength bands appear low in comparison to the total infrared flux, which is expected to scale with the bolometric luminosity, and the resulting SFRs from the 70 and 160~$\mu$m data are also lower.  Additionally, longer wavelength infrared bands are typically expected to include emission from diffuse dust heated by older stars \citep[e.g. ][]{bendo15a}, which should affect the SFRs based on the data from the bands \citep{calzetti10}.  However, if diffuse dust (particularly extended cold dust emission from outside the central starburst) was present, the SFRs should be biased upwards at longer wavelengths.  The low SFRs from the 70 and 160~$\mu$m data indicate that the 70 and 160~$\mu$m bands contain relatively little cold, diffuse dust, at least compared to the galaxies used by \citet{calzetti10} to derive the relations between SFR and emission in these bands.

Notably, the difference between the SFRs from the total infrared and H30$\alpha$ fluxes is less than 1$\sigma$ or 5\% before any corrections for metallicity or stellar rotation are applied.  This increases to 1.5-2$\sigma$ or 20-30\% when the correction are applied, but given the number of assumptions behind the corrections as well as the calculation of the SFRs themselves, this match is actually very good.  Aside from the potential calibration issues, the total infrared flux may also yield a slightly higher SFR because the dust emission could also include some energy absorbed from an older stellar population.  However, given the low SFRs from the 70 and 160~$\mu$m bands, which would be more strongly affected by an older stellar population, it seems very unlikely that any such diffuse dust emission, especially extended dust emission outside the central starburst, contributes significantly to the total infrared flux.  It is also possible that photoionizing photons are directly absorbed by dust grains before they are absorbed by the ionized gas, which could cause a slight difference in the SFRs.  None the less, calibration issues are probably one of the main reasons for any discrepancies between the SFRs from the H30$\alpha$ and total infrared fluxes.

\subsection{Optical and near-infrared recombination line data}

\subsubsection{SFR calculations}
\label{s_othersfr_hacalc}

\begin{table}
\caption{H$\alpha$ fluxes for the centre of NGC~5253$^a$.}
\label{t_ha}
\begin{center}
\begin{tabular}{@{}lc@{}}
\hline
Extinction Correction &
  Flux (erg cm$^{-2}$ s$^{-1}$ \\
\hline
No correction$^b$ &
  $(8.9 \times 2.0)\times10^{-13}$ \\
Correction by \citet{calzetti15} &
  $(1.7 \pm 0.5)\times10^{-11}$ \\
Correction using 24$\mu$m flux density$^c$ &
  $(3.47 \pm 0.19)\times10^{-11}$\\
Correction using total infrared flux$^c$ &
  $(8.6 \pm 0.5)\times10^{-12}$\\
\hline
\end{tabular}
\end{center}
$^a$ All H$\alpha$ fluxes are based on the sum of fluxes from clusters 5 and 13 from \citet{calzetti15}.\\
$^b$ These data include foreground dust extinction corrections but no corrections for intrinsic dust extinction.\\
$^c$ Infrared fluxes are based on the globally-integrated measurements, which primarily originate from a pointlike source; see Section~\ref{s_othersfr_ir} for details.\\
\end{table}

While the two central H$\alpha$ and Pa$\alpha$ sources are the brightest optical and near-infrared recombination line sources seen in this galaxy, multiple other fainter sources are also detected by \citet{alonsoherrero04}, \citet{harris04}, and \citet{calzetti15}, and diffuse extended emission is also present in the H$\alpha$ image from \citet{dale09}.  While a comparison could be made between the global H$\alpha$ and H30$\alpha$ emission, the inclusion of these fainter sources would cause some inaccuracies.  We therefore compared the SFR from the H30$\alpha$ data to the SFR from the near-infrared and optical line data from \citet{calzetti15}.  As indicated in Section~\ref{s_images}, it is possible that the two central clusters identified by \citet{alonsoherrero04} and \citet{calzetti15} are actually parts of one larger photoionization region, so we will use the sum of the optical and near-infrared fluxes from these two sources in our analysis.

\citet{calzetti15} published H$\alpha$, Pa$\alpha$, and Pa$\beta$ fluxes that have been corrected for foreground dust extinction but not for intrinsic dust extinction within the galaxy.  Using these data, they then derive the intrinsic dust extinction for the sources as well as extinction-corrected H$\alpha$ fluxes.  The corrections for cluster 5 assume a simple foreground dust screen, but cluster 11 is treated as a case where the line emission is attenuated both by a foreground dust screen and by dust intermixed with the stars.  The sum of these corrected H$\alpha$ fluxes, which are listed in Table~\ref{t_ha}, are used for computing one version of SFRs.

We also tested SFRs calculated by correcting H$\alpha$ for dust attenuation by adding together the observed H$\alpha$ emission (representing the unobscured emission) and infrared emission multiplied by a constant (representing the obscured H$\alpha$ emission).  Such constants have been derived for multiple individual infrared bands, including the {\it Spitzer} 8 and 24~$\mu$m bands and the WISE 12 and 22~$\mu$m bands, as well as the total infrared flux \citep[e.g. ][]{calzetti05, calzetti07, kennicutt07, zhu08, kennicutt09, lee13}.  To be concise, we restrict our analysis to the WISE 22~$\mu$m and total infrared flux.  The relations we used for these corrections are 
\begin{equation}
f(\mbox{H}\alpha)_{corr} = f(\mbox{H}\alpha)_{obs} 
  + 0.020 \nu f_\nu(22 \mu\mbox{m})
\label{e_sfr_ha_22}
\end{equation}
and
\begin{equation}
f(\mbox{H}\alpha)_{corr} = f(\mbox{H}\alpha)_{obs} 
  + 0.0024 f(\mbox{TIR}),
\label{e_sfr_ha_tir}
\end{equation}
which are from \citet{kennicutt09}.  Based on the dispersions in the data used to derive these relations, the uncertainty in the result from Equation~\ref{e_sfr_ha_22} and \ref{e_sfr_ha_tir} are 0.12 dex (32\%) and 0.089 dex (23\%), respectively.  The sum of the uncorrected H$\alpha$ fluxes for clusters 5 and 11 from \citet{calzetti15} as well as versions of these fluxes corrected using Equations \ref{e_sfr_ha_22} and \ref{e_sfr_ha_tir} are listed in Table~\ref{t_ha}.

The H$\alpha$ fluxes can be converted to SFR using
\begin{equation}
\frac{\mbox{SFR}(\mbox{H}\alpha)}{\mbox{M}_\odot\mbox{ yr}^{-1}}=6.43\times10^8
  \left[\frac{D}{\mbox{ Mpc}}\right]^{2}
  \left[\frac{f(\mbox{H}\alpha)}{\mbox{erg s}^{-1}\mbox{ cm}^{-2}}\right]
\label{e_sfr_ha}
\end{equation}
from \citet{murphy11}.  This result is based on models from {\sc Starburst99} using a Kroupa IMF, solar metallicities, and older but unspecified stellar evolution tracks.  The assumed $T_e$ is 10000~K, which is close to the measured value in NGC~5253, so we will make no modifications to this conversion. The SFRs can be rescaled to correct for metallicity and stellar rotation effects in the same way that the SFR from the H30$\alpha$ was rescaled.  The SFRs based on the three different extinction-corrected H$\alpha$ fluxes are listed in Table~\ref{t_sfr}.

\subsubsection{Discussion}

The extinction-corrected H$\alpha$ fluxes calculated by \citet{calzetti15} using Pa$\alpha$ and Pa$\beta$ line data yield SFRs that fall within 25\% of the SFRs from the H30$\alpha$ data.  Given that the extinction corrections changed the H$\alpha$ fluxes by $\sim$20$\times$, that relatively complex dust geometries were used in calculating the corrections, and that the uncertainties in the extinction-corrected H$\alpha$ fluxes are relatively high, this match is reasonably good.  However, the fact that the SFR from the H30$\alpha$ is higher would indicate that the method of correcting the H$\alpha$ flux could still be improved.

We discussed in Section~\ref{s_images} the non-detection of a second H30$\alpha$ source corresponding to the secondary star forming region seen in optical and near-infrared bands, which is labelled as cluster 5 by \citet{calzetti15}.  One possible explanation for this is that the optical/near-infrared sources lie at the ends of a much larger star forming complex that is heavily obscured in its centre.  If this is the case, then the ratio of the area of the optical/near-infrared sources to the area of the much larger star forming complex should be roughly similar to the ratio of the SFR from the extinction-corrected H$\alpha$ emission to the SFR from the H30$\alpha$ emission.  \citet{calzetti15} do not list sizes for their sources, but they do state that they use measurement apertures with diameters of 0.25~arcsec, which we can treat as an upper limit on the source sizes.  The hypothetical larger star forming complex would need to be 0.71~arcsec in size to encompass both optical/near-infrared regions.  Based on the ratio of the area of the two smaller optical/near-infrared sources to the area of the hypothetical larger complex (which we can assume is spherical), the optical/near-infrared regions should yield a SFR that is only 25\% of that from the H30$\alpha$ emission, not the 75\% that we measure.  This indicates that it is unlikely that both optical/near-infrared sources are part of one larger obscured complex centered on the H30$\alpha$ source.

The other possible reason for the non-detection of a second H30$\alpha$ source corresponding to cluster 5 from \citet{calzetti15} is that the difference in brightness between it and the brighter source (cluster 11) is higher than the factor of $\sim$6$\times$ derived from the analysis of the optical and near-infrared data.  The results here indicate that the extinction correction for the H$\alpha$ data applied by \citet{calzetti15} may indeed be too low, so it is possible that the difference in the extinction-corrected brightness between the two sources is much greater than implied by their analysis.  In such a scenario, the second source would be too faint in recombination line emission to detect in the ALMA data at the 3$\sigma$ level.

The SFR based on the composite H$\alpha$ and 22~$\mu$m data yields a SFR that is highest among the three based on the H$\alpha$ data, and it is also significantly higher than the SFR from the H30$\alpha$ flux.  This is most likely a result of the unusually hot dust within NGC~5253, and the problem is similar to the direct conversion of 22~$\mu$m flux density to SFR discussed in Section~\ref{s_othersfr_ir}.  The extinction correction in Equation~\ref{e_sfr_ha_22} was calibrated using spiral galaxies and, to some degree, relied on a linear relation between the amount of energy absorbed by dust and mid-infrared emission.  Since that scaling relation does not apply well to NGC~5253, Equation~\ref{e_sfr_ha_22} overcorrected the extinction.

In contrast, the composite H$\alpha$ and total infrared flux yielded a SFR that was too low compared to the H30$\alpha$ value or most other values listed in Table~\ref{t_sfr}.  Again, this is probably because the relation was calibrated using spiral galaxies.  The emission from cold, diffuse dust heated by evolved stars seems to be a relatively small fraction of the total infrared emission in NGC~5253 in comparison to what is found in spiral galaxies.  If Equation~\ref{e_sfr_ha_tir} is effectively calibrated to account for this cold dust, then it may yield an underestimate of the extinction correction in galaxies like NGC~5253.  Also note that any attempt to correct the total infrared flux for extended emission outside the central starburst will simply make the SFR from Equation~\ref{e_sfr_ha_tir} more discrepant comapred to values calculated from the H30$\alpha$ data or most other bands.

The relative sizes of the uncorrected H$\alpha$ and infrared components in Equations~\ref{e_sfr_ha_22} and \ref{e_sfr_ha_tir} provide some additional insights into these extinction corrections.  In both cases, the infrared component is dominant.  In Equation~\ref{e_sfr_ha_22}, the observed H$\alpha$ flux is equivalent to 2.6\% of the 22~$\mu$m term, so using Equation~\ref{e_sfr_ha_22} and \ref{e_sfr_ha} to calculate a SFR is effectively an indirect conversion of the 22~$\mu$m flux density to SFR.  Meanwhile, the uncorrected H$\alpha$ flux in Equation~\ref{e_sfr_ha_tir} is equal to 11.5\% of the total infrared flux term, so using the resulting corrected H$\alpha$ flux in Equation~\ref{e_sfr_ha} is not quite as much like indirectly converting the total infrared flux to SFR.

Related to this, \citet{kennicutt12} and references therein describe how infrared flux could be used correct ultraviolet flux densities in the same way as they correct H$\alpha$ fluxes.  However, when we investigated using such equations with the ultraviolet flux densities measured for the central two star clusters, we found that the infrared terms were $>$100$\times$ higher than the ultraviolet terms, which meant that any SFR based on combining ultraviolet and infrared data would effectively be independent of the ultraviolet measurements.

\section{Conclusions}
\label{s_conclu}

To summarize, we have used ALMA observations of H30$\alpha$ emission from NGC~5253, a low-metallicity blue compact dwarf galaxy, in a comparison with different methods of calculating SFR for the centre of this galaxy.  We measure a $Q$ (1.9$\pm$0.3)$\times$$10^{52}$~s$^{-1}$, which is based on using a distance of 3.15~Mpc.  Accounting for the low metallicity of the galaxy and stellar rotation, we obtain an SFR of 0.087$\pm$0.013 M$_\odot$ yr$^{-1}$; with only the correction for metallicity, we obtain 0.102$\pm$0.015 M$_\odot$ yr$^{-1}$.

In our analysis, we found three SFR measurements that best matched the H30$\alpha$ measurements and that seemed to be the least affected by the types of systematic effects that we could identify as causing problems with other bands.

The total infrared flux (calculated by integrating the SED between 12 and 500~$\mu$m) yielded a SFR that was very similar to the value from the H30$\alpha$ data.  In other dusty low-metallicity starbursts like NGC~5253, the total infrared flux may yield the most accurate star formation rates as long care is taken to account for the unusual shapes of the SEDs for these galaxies.

The 70~$\mu$m band may be the best monochromatic infrared star formation tracer available, as it yielded a SFR that was closest to the SFRs from either the H30$\alpha$ and total infrared flux.  However, given how the conversion of 70~$\mu$m flux to SFR depends on a linear relationship between emission in this band and the total infrared flux, it is not clear how reliable 70~$\mu$m emission would be for measuring SFR in other low metallicity galaxies where a relatively large but potentially variable fraction of the dust emission is at mid-infrared wavelengths.

The SFR from the H$\alpha$ flux that was extinction corrected by \citet{calzetti15} using Pa$\alpha$ and Pa$\beta$ data was 25\% lower than but consistent with the SFR from the H30$\alpha$ data.  However, as we noted in Section~\ref{s_images}, it is possible that some parts of the central star forming complex are completely obscured in the optical and near-infrared observations, which potentially illustrates the issues with examining star formation within dust starbursts, even at near-infrared wavelengths.

Most other star formation tracers that we examined seemed to be affected by systematic effects that cause problems when calculating SFRs.

Previously-published versions of the SFR based on millimetre and radio data yielded star formation rates with a lot of scatter relative to each other and relative to the SFR from the H30$\alpha$ data.  At least some of the SFRs from radio continuum measurements are affected by incorrect assumptions about the nature of the emission sources in these bands.  A new analysis of broadband archival radio and millimetre data is needed to produce better models of the SED and to covert emission from these bands into SFRs more accurately.

The 22~$\mu$m flux density by itself and the combined H$\alpha$ and 22~$\mu$m metric produced SFRs that were exceptionally high compared to the value from the H30$\alpha$ data and compared to SFRs from other tracers.  The main problem is that the dust in this low metallicity environment is thinner than in solar metallicity objects, so the dust that is present is exposed to a brighter and hotter radiation field.  Consequently, the total dust emission is stronger in low metallicity environments, and the mid-infrared emission is stronger relative to total infrared emission.  Based on these results, we strongly recommend not using any star formation tracer based on mid-infrared data for low metallicity galaxies.

Infrared emission at 160~$\mu$m yields a very low SFR.  This is probably because the dust temperatures within NGC~5253 are relatively hot and because the conversion of emission in this band to SFR accounts for the presence of diffuse dust heated by older stars that is present in spiral galaxies but not in NGC~5253 and similar dwarf galaxies.

The composite H$\alpha$ and total infrared star formation metric yielded a SFR that was too low.  Again, this could be because the correction is calibrated using spiral galaxies that contain relatively higher fractions of diffuse, cold dust than NGC~5253.  If such a metric were to be used for measuring star formation in low metallicity systems, it would need to be recalibrated.

This analysis represents the first results from using ALMA observations of hydrogen millimetre recombination line emission to test SFR metrics based on optical data, and it is also one of the first comparisons of SFR metrics for a low metallicity galaxy that has involved ALMA data.  A more thorough analysis of star formation tracers observed in low metallicity galaxies is needed to understand whether the results obtained for NGC~5253 are generally applicable to similar objects.  Further analysis of more ALMA recombination line observations combined with this and previous works on this subject \citep{bendo15b, bendo16} will allow us to create a broader picture of the reliability of various tracers of star formation in both nearby starbursts and more distant galaxies.

\section*{Acknowledgments}

We thank the reviewer for the constructive criticisms of this paper.  GJB and GAF acknowledge support from STFC Grant ST/P000827/1.  KN acknowledges support from JSPS KAKENHI Grant Number 15K05035.  CD acknowledges funding from an ERC Starting Consolidator Grant (no.~307209) under FP7 and an STFC Consolidated Grant (ST/L000768/1).  This paper makes use of the following ALMA data: ADS/JAO.ALMA\#2013.1.00210.S.  ALMA is a partnership of ESO (representing its member states), NSF (USA) and NINS (Japan), together with NRC (Canada), NSC and ASIAA (Taiwan), and KASI (Republic of Korea), in cooperation with the Republic of Chile. The Joint ALMA Observatory is operated by ESO, AUI/NRAO and NAOJ.  This research has made use of the NASA/ IPAC Infrared Science Archive, which is operated by the Jet Propulsion Laboratory, California Institute of Technology, under contract with the National Aeronautics and Space Administration.  This publication makes use of data products from the Wide-field Infrared Survey Explorer, which is a joint project of the University of California, Los Angeles, and the Jet Propulsion Laboratory/California Institute of Technology, and NEOWISE, which is a project of the Jet Propulsion Laboratory/California Institute of Technology. WISE and NEOWISE are funded by the National Aeronautics and Space Administration.

{}

\appendix

\label{lastpage}

\end{document}